\documentclass[pdflatex,sn-nature]{sn-jnl}

\usepackage{graphicx}%
\usepackage{multirow}%
\usepackage{amsmath,amssymb,amsfonts}%
\usepackage{amsthm}%
\usepackage{mathrsfs}%
\usepackage[title]{appendix}%
\usepackage{xcolor}%
\usepackage{textcomp}%
\usepackage{manyfoot}%
\usepackage{booktabs}%
\usepackage{algorithm}%
\usepackage{algorithmicx}%
\usepackage{algpseudocode}%
\usepackage{listings}%
\usepackage[left]{lineno}

\theoremstyle{plain}%
\theoremstyle{thmstyletwo}%

\theoremstyle{thmstylethree}%

\begin{document}
\title[Reentrant Landau Levels in a Dirac topological insulator]{Reentrant Landau Levels in a Dirac topological insulator}

\author[1,2]{C.~Kaufmann Ribeiro}
\author[3]{J.~C.~Mutch}
\author[3]{Q.~Jiang}
\author[3]{J.~P.~Ayres-Sims}
\author[1]{K.~Rubi}
\author[1]{C.~A.~Mizzi}
\author[1]{E.~A.~Peterson}
\author[4]{D.~Bulmash}
\author[1]{J.~Singleton}
\author[1]{N.~Harrison}
\author[1]{P.~F.~S.~Rosa}
\author[1]{J.-X.~Zhu}
\author[3]{J.-H.~Chu}
\author[2]{J.~Larrea Jiménez}
\author[1]{S.~M.~Thomas}
\author[1]{J.~C.~Palmstrom}

\affil[1]{Los Alamos National Laboratory, Los Alamos, NM 87545, USA}
\affil[2]{Laboratory for Quantum Matter under Extreme Conditions, Institute of Physics, University of São Paulo, São Paulo, Brazil}
\affil[3]{Department of Physics, University of Washington, Seattle, WA 98195, USA}
\affil[4]{Department of Physics, United States Naval Academy, Annapolis, MD 21402, USA}

\footnotetext{
Work of C.K.R. performed at Los Alamos National Laboratory, Los Alamos, NM 87545, USA; 
Home institution Laboratory for Quantum Matter under Extreme Conditions, 
Institute of Physics, University of São Paulo, São Paulo, Brazil.
}

\abstract{
The quantum limit, where magnetic fields confine carriers to the lowest Landau level, 
is predicted to host exotic quantum phases arising from strengthened electronic correlations, 
reduced dimensionality, and increased degeneracy. We report a novel quantization regime realized in the ultra-quantum limit of the narrow-gap Dirac insulator ZrTe$_5$, marked by anomalous magnetoresistance oscillations. These oscillations, measured in ZrTe$_5$ single crystals down to 700~mK and up to 60~T, are distinctly non-\(1/B\) periodic and persist for magnetic fields well beyond the quantum limit. In this regime, the competition between Zeeman and cyclotron energies drives a nonlinear evolution and back-bending of Landau levels, causing low-index levels to re-cross the Fermi energy at high fields. This mechanism departs from the standard Lifshitz--Kosevich description and provides a framework to describe how the electronic structure in topological Dirac insulators evolves beyond the quantum limit.
}




\maketitle
     \pagebreak

\section{Introduction}\label{sec1}

Accessing the quantum limit (QL)  is particularly compelling because, in this extreme regime where electrons are confined to their lowest Landau level (LL), systems become highly susceptible to electronic instabilities arising from reduced dimensionality and enhanced degeneracy~\cite{shoenberg2009magnetic, tsui1982two, halperin1987possible,galeski2021origin}. These instabilities can generate a rich landscape of emergent phases, including excitonic insulators, quantum Hall states, and spin or charge density waves~\cite{fenton1968excitonic,Khveshchenko2001,LuisA,FradkinKivelson1999,klitzing1980new,kohmoto1992diophantine,bernevig2007theory,halperin1987possible,Jain2007Composite}. However, reaching the QL typically requires applying extremely strong magnetic fields, often far beyond standard experimental capabilities. In the case of 3D electron systems, only a few semimetals with exceptionally low carrier densities and small effective masses - such as TaAs \cite{ramshaw2018quantum}, NbP \cite{modic2019thermodynamic}, Bi \cite{Zhu2017}, graphite \cite{iye1982high,zhu2019graphite}, and notably ZrTe$_5$ \cite{liu2016zeeman,tang2019three,Zhang2020,galeski2021origin,gourgout2022magnetic,wang2021magneto} - offer the rare opportunity to access the QL in experimentally achievable fields.

ZrTe$_5$ is a van der Waals (vdW) layered material, with an orthorhombic crystal structure and quasi-2D ZrTe$_3$ chains running along the $a$ axis and stacked along the $b$ axis  (Figure~\ref{ZrTe5_zero_field}a). ZrTe$_5$ hosts a nontrivial band topology driven by a band inversion associated with its nonsymmorphic space group symmetry (C$_\textit{mcm}$  (D$_{2h}$) ) \cite{weng2014transition}. Sitting near the phase boundary between weak and strong topological insulators \cite{weng2014transition, fan2017transition, manzoni2016evidence, mutch2019evidence}, ZrTe$_5$ is effectively described by a 3D Dirac Hamiltonian, with  Dirac fermions exhibiting ultralow carrier densities, small effective masses, and ultrahigh mobilities \cite{galeski2022signature, jiang2017landau, tang2019three, wang2018vanishing}. These properties not only allow the system to access the QL under moderate magnetic fields (below $\approx 3\,\mathrm{T}$) but also give rise to a rich variety of emergent phenomena, such as negative longitudinal magnetoresistance  associated with the magnetic chiral anomaly \cite{li2016chiral}, a temperature-induced metal-insulator transition linked to a Lifshitz transition reshaping its Fermi surface topology \cite{chi2017lifshitz,gourgout2022magnetic,shahi2018bipolar, zhang2017electronic}, and exceptionally large magnetoresistance (MR) across wide temperature and field ranges \cite{liu2016zeeman,galeski2022signature,wang2021magneto}.

Among the most intriguing discoveries in ZrTe$_5$ is the observation of unconventional MR oscillations that deviate from the standard \(1/B\)-periodic behavior, persisting above the QL and extending to high magnetic fields \cite{wang2018discovery,xing2024rashba}. These oscillations have challenged the conventional understanding of quantum oscillations (QO), which, in the vast majority of materials, are governed by the semiclassical Landau quantization condition \cite{shoenberg2009magnetic}. In this standard framework, QO arise from the successive crossing of LLs through the Fermi energy, producing a  \(1/B\)-periodic modulation that is expected to vanish once the system reaches the QL \cite{shoenberg2009magnetic}. Interestingly, the oscillatory behavior in ZrTe$_5$ is strongly sample dependent: while some samples exhibit anomalous non-\(1/B\)-periodic oscillations, others show conventional \(1/B\)-periodic oscillations consistent with the standard Lifshitz-Kosevich (LK) framework \cite{galeski2021origin, galeski2022signature, wang2018vanishing}.  This variability raises questions about the underlying physical mechanisms governing the oscillations and the ground state electronic structure of ZrTe$_5$. 

In this work we investigated the high-field regime of ZrTe$_5$, well beyond the QL. Our MR experiments, performed at temperatures as low as 700 mK and in magnetic fields up to 60 T, revealed  non-$1/B$ periodic oscillations that persist to the highest available  fields. We also observed non-monotonic temperature dependence in the oscillatory amplitudes, indicating a departure from the conventional LK framework. Furthermore, our data suggests that the electronic effective mass increases with the magnetic field. Finally,  angular magnetoresistance (AMR) measurements tracking the MR oscillations, indicate a 3D ellipsoidal Fermi surface at low magnetic fields. Non $1/B$-periodic oscillations have been previously observed in several studies, both in the high field regime (up to 60 T) where their presence has been attributed to a manifestation of discrete scale invariance  \cite{wang2018discovery} and the intermediate field regime (up to 9 T) where the Rashba effect has been proposed as a possible explanation \cite{xing2024rashba}. In this work we propose a unifying model spanning all field regimes.

To explain our data we employ well-established  low-energy \(k \cdot p\) model that captures the nonlinear LL dispersion arising from the competition between cyclotron and Zeeman energies in the presence of  strong spin-orbit coupling (SOC) \cite{chen2015magnetoinfrared, galeski2022signature, wang2021magneto, jiang2017landau}.  This interplay causes LL "back-bending", allowing low-index levels to re-cross the Fermi energy above the QL. This model successfully accounts for the anomalous observations in our experiments, including the non-$1/B$ periodic oscillations, non-LK temperature dependence, and AMR. Importantly, we show that this framework naturally reconciles both the conventional $1/B$-periodic and all anomalous non-$1/B$-periodic oscillations reported in ZrTe$_5$, providing a unified description. This offers an important step toward understanding the role of sample dependence in topological Dirac insulators and semimetals.

\section{Results and discussion}

\subsection{Semiconductor-like behavior}

\begin{figure}[H]
	\includegraphics[width=1\textwidth]{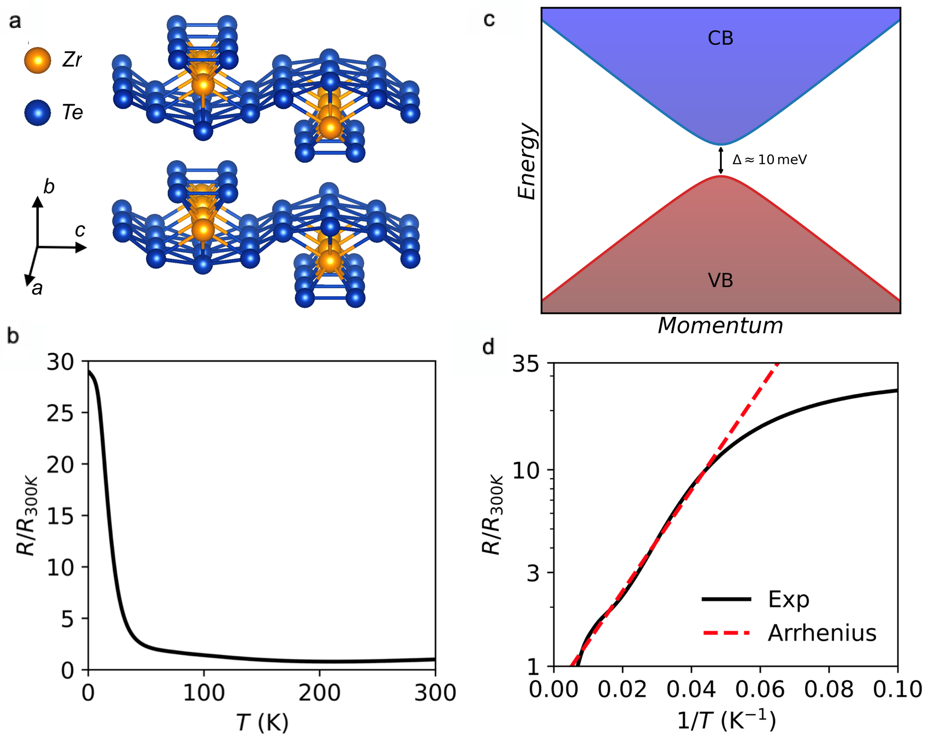}
	\centering
	\caption{ (a) Schematic of the crystal structure of ZrTe$_5$, showing chains of ZrTe$_3$ prisms aligned along the $a$ axis. These chains are interconnected along the $c$ axis via Te atoms, forming 2D layers that are stacked along the $b$ axis and held together by vdW interactions. (b) Electrical resistance of flux grown ZrTe$_5$.  Resistance was measured along the $a$ axis, normalized by the room temperature resistance ($R_{300\,\text{K}}$). The sharp resistance increase at low temperature is consistent with semiconducting behavior.  (c) Schematic representation of Dirac bands in ZrTe$_5$ with a gap ($\Delta$) of 10 meV, illustrating the electronic structure near the Dirac points. (d) $\log$-linear plot of $R/R_{300\text{ K}}$  vs $1/T$  of the resistance measurement shown in (b) (black line). Overlaid is an Arrhenius fit in the temperature range of 20 K-150 K (red dashed line)  with fitted activation energy of 5 meV. Details of the fitting procedure are discussed in the main text.}
	\label{ZrTe5_zero_field}
\end{figure}

Figure~\ref{ZrTe5_zero_field}b shows the temperature dependence of the electrical resistance, $R(T)$, of the flux grown \cite{mutch2019evidence}, single crystal ZrTe$_5$ sample  studied in this work. The resistance exhibits semiconducting behavior from 200~K down to 0.4~K, with a clear saturation below 10~K (see  Supplementary Materials). This behavior is consistent with previous reports on flux-grown ZrTe$_5$ crystals~\cite{mutch2019evidence,wang2018discovery,shahi2018bipolar,xing2024rashba}. However, it is important to note that even among flux-grown samples, significant variability exists. The energy gap and the position of the Fermi energy can differ markedly from sample to sample, with reported gap values ranging from as low as 10~meV to as high as 60~meV \cite{mutch2019evidence,li2016experimental,Chen2017,jiang2017landau,xiong2017three}. These variations can be attributed to sample-dependent strain \cite{mutch2019evidence} and the presence of tellurium vacancies \cite{peterson2024te}, all of which can shift band edges and modulate the carrier density. In contrast, chemical vapor transport (CVT)-grown samples typically display a pronounced peak in $R(T)$ at intermediate temperatures \cite{okada1980giant}, commonly attributed to a Lifshitz transition involving a temperature-induced shift of the Fermi energy across a band extremum \cite{zhang2017electronic,chi2017lifshitz}.

Assuming that the dominant mechanism governing the temperature dependence of the resistance is thermally activated excitation of carriers across a band gap (as illustrated in Figure~\ref{ZrTe5_zero_field}c), we fit the resistance data in the 20–150 K range using the Arrhenius model, $R(T) = R_1 \exp(E_a / k_B T)$. Here $R_1$ is a prefactor, $E_a$ is the activation energy, $k_B$ is the Boltzmann constant, and $T$ is the temperature. This approach captures the exponential dependence of resistance on inverse temperature, characteristic of gapped semiconductors. \footnote{Angle-resolved photoemission spectroscopy (ARPES) studies have reported that the Fermi energy in ZrTe$_5$ can vary significantly with temperature~\cite{zhang2017electronic}, often moving relative to the band edges due to its proximity to the conduction and valence bands. While such shifts could, in principle, influence the effective activation energy extracted from transport measurements, the good agreement of the data with the Arrhenius model in the temperature window [20 K,150 K] suggests that to first order these contributions can be neglected.}

As shown in Figure \ref{ZrTe5_zero_field}d, the Arrhenius fit yields an activation energy of $E_a = 5$ meV, implying a band gap of approximately $\Delta \approx 10$ meV under the assumption $E_a \approx \Delta/2$. This value reflects the minimum energy required for thermal excitation of carriers into conducting states and supports the interpretation of semiconducting behavior at intermediate temperatures. A more detailed analysis of the temperature dependence of the electrical resistivity, using a Dirac dispersion model, is presented in the Supplementary Material.

\subsection{Non-1/B Periodic MR oscillations}

\begin{figure}[H]
	\includegraphics[width=1\textwidth]{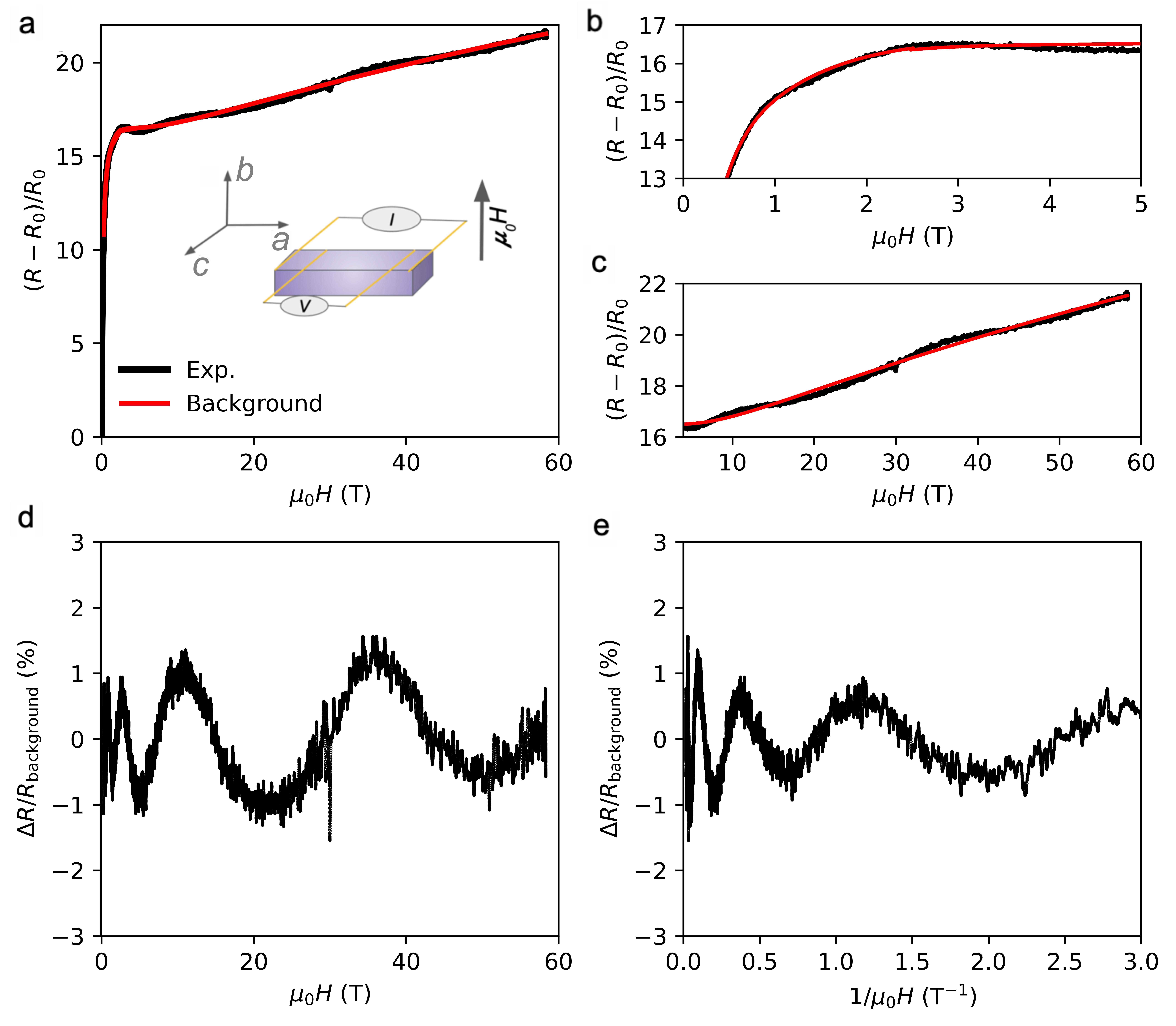}
	\centering
	\caption{Magnetoresistance  of ZrTe$_5$ with the current applied along the $a$ axis and the magnetic field applied along the $b$ axis at 4 K. (a) Magnetoresistance data up to 60 T, where the black curve represents the experimental data and the red curve indicates the background fit. The inset shows a schematic representation of the experimental configuration. Zoom in of the low field (b) and high field (c) regions of the magnetoresistance. (d)  $\Delta R/R_{background}$ as a function of magnetic field to show the evolution of the magnetoresistance oscillations in field. (e) Same data shown in (b) plotted as a function of inverse magnetic field, highlighting the deviation from $1/B$ periodicity.}
	\label{Pulsed_MR}
\end{figure}

Figure~\ref{Pulsed_MR}a shows the MR of ZrTe$_5$ at 4~K in a pulsed magnetic field up to 60~T, normalized by $R_0=R(B=0)$, with current applied along the crystallographic $a$ axis and magnetic field along the $b$ axis ($\mu_0H \parallel b$). The MR increases rapidly below 2~T, exceeding 2000\% at 60~T. This exceptionally large MR is consistent with previous reports on both flux-grown and CVT-grown ZrTe$_5$ samples~\cite{tritt1999large,li2016chiral,shahi2018bipolar}.

In addition to the rapid increase in MR, oscillations are observed as a function of the magnetic field from 0.3~T to 60~T. To analyze the oscillatory behavior, we performed a background ($R_{background}$) subtraction (see Supplementary Materials) where $\Delta R=R(B)-R_{background}(B)$. The oscillations are clearly visible throughout the entire measurement range, as shown in Figures~\ref{Pulsed_MR}b and \ref{Pulsed_MR}c. The low-field (below 0.3 T) region is difficult to analyze, as any potential MR oscillations are likely merged into the sharp increase in MR near zero field, a limitation also noted in previous studies  \cite{wang2018discovery}.

The persistence of oscillations at high magnetic fields (Figure \ref{Pulsed_MR}) represents a behavior that is not universally observed in ZrTe$_5$ \cite{liu2016zeeman,wang2018vanishing,tang2019three,galeski2021origin,galeski2022signature} and is inconsistent with the standard understanding of SdH oscillations~\cite{liu2016zeeman,PhysRevB.93.115414,shoenberg2009magnetic}. Given the very low effective mass and high Fermi velocity in this material, for low carrier density samples the QL is expected to occur below 3~T~\cite{galeski2021origin, galeski2022signature, wang2018discovery}. In the QL, all carriers are confined to the lowest LL, and conventional SdH oscillations cease. However, in our measurements, the MR oscillations continue well beyond this point, suggesting the presence of unconventional mechanisms. 

After a polynomial background subtraction and normalization (see Supplementary Materials), the MR oscillations are found to deviate from the standard $1/B$-periodic behavior expected from SdH oscillations. As shown in Figure~\ref{Pulsed_MR}e, the oscillation extrema are not evenly spaced in inverse magnetic field. Similar behavior has been observed in other ZrTe$_5$ samples, including MR oscillations above the QL ~\cite{wang2018discovery,xing2024rashba} (see Supplementary Materials). It is worth-noting that the observed oscillations show the same apparent $\log(B)$ periodicity of previous reported measurements \cite{wang2018discovery}  in the available field range (up to 60 T) (see Figure \ref{Landau_levels}a).

\subsection{Role of Zeeman energy in Landau Levels in Dirac systems}

We now examine how the relativistic nature of Dirac fermions in ZrTe\(_5\) influences the LL spectrum under high magnetic fields. In particular, we analyze the competition between Zeeman and cyclotron energies and how their interplay affects the behavior of oscillations in a narrow-gap Dirac semimetal.

In systems with large SOC, such as ZrTe\(_5\), Zeeman splitting can no longer be simply treated as an additive term to a spin-independent band structure. The SOC entangles spin and momentum, so the Hilbert space cannot be decomposed into spin and orbital sectors. This leads to a more complex field dependence of the LL energies. In such systems, the effective \(g\)-factor can be extremely large, and the Zeeman contribution must be explicitly included when describing the LL structure.

To quantify these effects, we adopt a low-energy \(k \cdot p\) Hamiltonian for a 3D Dirac system with time-reversal, inversion, and mirror symmetries. This has previously been used to fit the  field dependence in ZrTe$_5$~\cite{chen2015magnetoinfrared, galeski2022signature, wang2021magneto, jiang2017landau}:
\begin{equation}
    H(k)=\hbar(v_a k_a \tau^a \sigma^b + v_c k_c \tau^c + v_b k_b \tau^a \sigma^a) + M \tau^b - \frac{\mu_B g \sigma^b}{2} B
\end{equation}

Here, \(v_{a,b,c}\) are Fermi velocities, \(k_{a,b,c}\) the momentum operators, \(\tau^{a,b,c}\) and \(\sigma^{a,b,c}\) are Pauli matrices in orbital and spin space, respectively, and \(M\) is the Dirac mass (\(\Delta = 2M\)). Here  the $a,b,c$ subscripts refer to the crystallographic directions. The Zeeman term is included for a magnetic field along the \(b\) axis .

\begin{figure}[H]
	\includegraphics[width=1\textwidth]{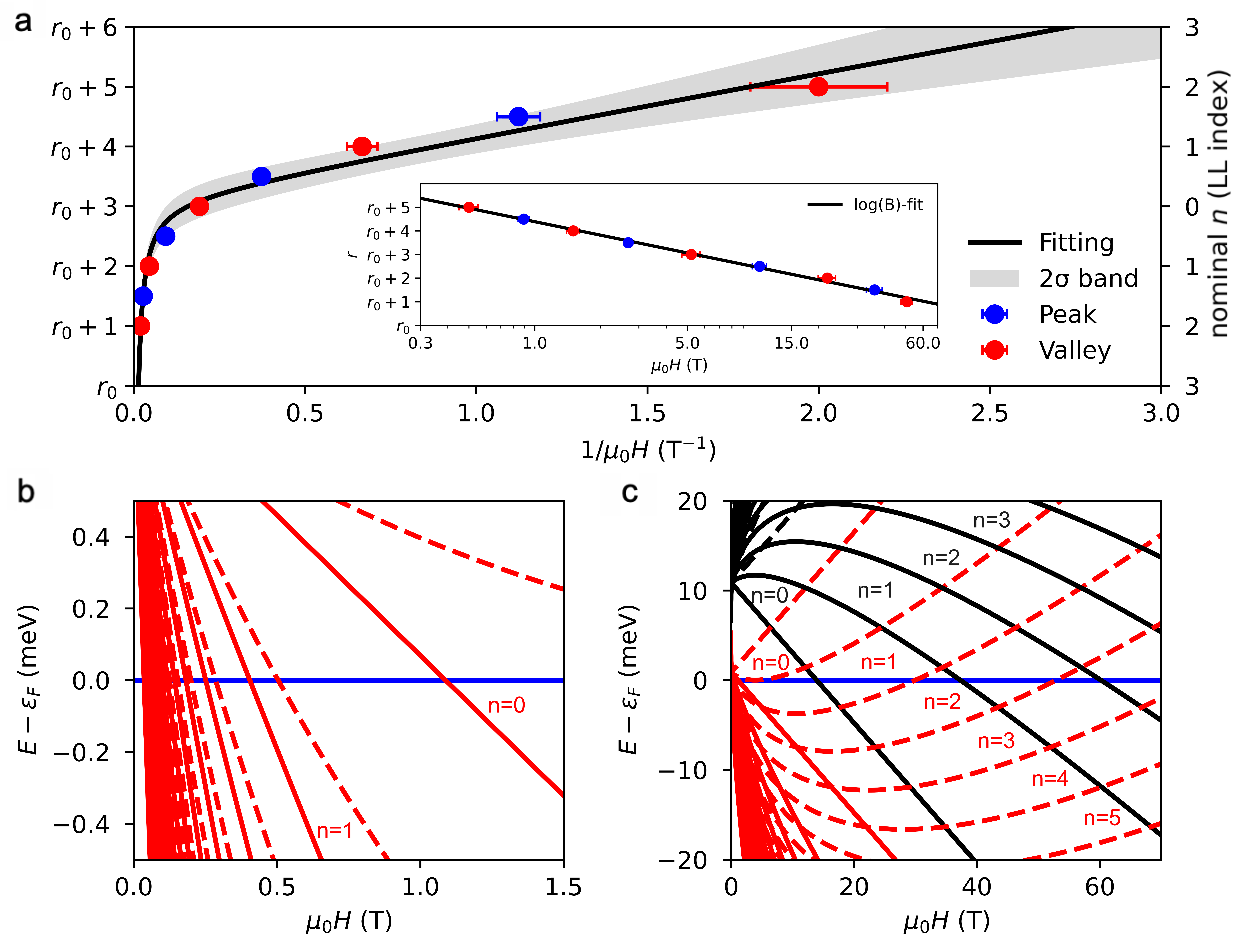}
	\centering
	\caption{(a) Magnetic-field positions of oscillation peaks (blue) and valleys (red), fitted using Eq.~\ref{r-fitting} (black line). With increasing field, the oscillation index $r$  increases. To demonstrate the re-entrant LL crossings we have arbitrarily picked $r_0=-3$ to assign a nominal LL index (right axis) to each point. The gray region indicates the validity of the model within 2$\sigma$ .The inset reveals an apparent log($B$)-periodicity resulting from the restricted field range. A fit of the oscillation index to $r(B) \approx A\log B + C$ gives $A = -0.82(2)$ and $C = 4.32(4)$.  (b and c) Calculated Landau level spectrum using the parameters and model discussed in the main text, shown over two magnetic-field windows for clarity. Solid lines represent spin-up states, while dashed lines denote spin-down states. Black curves correspond to conduction band LLs and red curves to valence band LLs. The Fermi energy is indicated by the blue line.}
    \label{Landau_levels}
\end{figure}

The resulting LL energies at the high-symmetry point are \cite{jiang2017landau,chen2015magnetoinfrared}:
\begin{equation}
E_{n,\lambda}^{s} = -s\frac{\mu_B g}{2} B + \lambda \sqrt{2 e \hbar v^2 n B + M^2},
\label{eigen_states}
\end{equation}
\(s = \pm1\) denotes spin-up and spin-down, \(\lambda = \pm1\) labels conduction and valence bands, \(v\) is an effective  velocity\footnote{Here $v=\sqrt{v_a v_c}$ is the in-plane Dirac velocity, with $v_a$ and $v_c$ representing the Dirac velocity along the $a$- and $c$-axes, respectivelly.  The effective Hamiltonian is constructed from a four-state $k\!\cdot\!p$ basis at the $\Gamma$ point, truncated to linear order and constrained by crystal symmetries.  A magnetic field is applied along the $b$  ($B \parallel b$), included via minimal coupling in Landau gauge. The Landau spectrum  follows from solving this Hamiltonian at $k_b=0$ using harmonic-oscillator eigenstates \cite{chen2015magnetoinfrared}.} , and $n$ is the LL index.

The LL energies given in Eq.~\ref{eigen_states} consist of two key contributions. The first term, $-s \frac{\mu_B g}{2} B$, corresponds to the Zeeman energy, which introduces a spin splitting that grows linearly with magnetic field. The second term, $\lambda \sqrt{2 e \hbar v^2 n B + M^2}$, reflects the cyclotron quantization of Dirac fermions, where the $\sqrt{B}$ dependence of the energy levels arises from the linear band dispersion. This behavior contrasts with the $B$-linear Landau quantization observed in systems with parabolic bands. Together, the interplay between the nonlinear cyclotron term and the linear Zeeman splitting leads to curvature in the LL evolution with field. At sufficiently high fields, this can result in the back-bending of LLs and their recrossing of the Fermi energy. The behavior is related to the reverse quantum limit (RQL), a model proposed to explain QOs in the insulating phase of YbB$_{12}$ \cite{mizzi2024reverse}. However, the underlying physics of ZrTe$_5$ and YbB$_{12}$, are distinct. ZrTe$_5$ can be described by a non-interacting Dirac Hamiltonian, whereas YbB$_{12}$, is a Kondo insulator.

To explore the impact on MR oscillations we consider the band crossing points condition, i.e., when \(E_{n,\lambda}^s = E_{\text{F}}\) and the frequency in which the LL cross the Fermi energy in $1/B$,  \(F(B) = \partial n_{E_{\text{F}}} / \partial (1/B)\), with $n_{E_{\text{F}}}$ defined as the n-index when a LL crosses the Fermi energy \cite{mizzi2024reverse,shoenberg2009magnetic}, yielding:
\begin{equation}
F(B) = \frac{E_{\text{F}}^2 - M^2}{2 e \hbar v^2} - \frac{\mu_B^2 g^2}{8 e \hbar v^2} B^2.
\label{Frequentcy}
\end{equation}

The first term is the $1/B$ contribution from the cyclotron energy and the second term is a correction arising from Zeeman coupling. Here we have assumed a constant g-factor as a function of magnetic field and we neglect the electronic dispersion parallel to the magnetic field. Due to the large \(g\)-factor in ZrTe\(_5\), this quadratic \(B\) term can significantly alter the oscillation frequency, especially at high fields and low \(E_{\text{F}}\), as detailed in the Supplementary Materials. Moreover, the frequency \(F(B)\) is spin- and band-independent, meaning that multiple LLs—e.g., spin-up and spin-down—can cross the Fermi energy at the same \(F\) but with phase shifts. This provides a natural explanation for the interference effects observed in the temperature dependence of MR oscillation amplitudes, discussed in the next section.

To test this model, we track the positions of peaks and valleys in the oscillatory MR. We then apply the LK framework, using:
\begin{equation}
    \Delta R / R_{\text{background}} = A(B,T) \cos\left[2\pi(F/B + \theta)\right]
    \label{LK_formula}
\end{equation} where \(A(B,T)\) is the oscillation amplitude\footnote{For arbitrary field evolutions of $F(B)$ and $\theta(B)$,  $A(B,T)$ would act as an amplitude envelope of the oscillatory behavior.  However, for the standard LK framework and the model considered here, A(B,T) is an oscillation amplitude.} and \(\theta\) is a phase shift. Each extremum satisfies \(|F(B)/B + \theta| = r\), where \(r\) is an integer for a peak and a half-integer for a valley.  Inserting into Eq.~\ref{Frequentcy} gives\footnote{Two important considerations are made: (1) the peak in the density of states occurs when a Landau level crosses the Fermi energy, i.e we assume a fixed chemical potential, which to first order is a good approximation; and (2) the magnetization of ZrTe$_5$ is small enough that $\mu_0 H \approx B$.}:
\begin{equation}
r(B) = \frac{E_{\text{F}}^2 - M^2}{2 e \hbar v^2} \cdot \frac{1}{B} - \frac{\mu_B^2 g^2}{8 e \hbar v^2} B + \theta.
\label{r-fitting}
\end{equation}

We emphasize that $r$ denotes the oscillation index (not the LLs index), so the absolute value of $\theta$ cannot be uniquely determined. Since the index of the first oscillation is not experimentally accessible, we reference all oscillations relative to an arbitrary index $r_0$. 

We set $M$ and $g$  using experimentally motivated values, with \(M = 5\) meV based on Arrhenius fits to thermally activated transport (see  Figure \ref{ZrTe5_zero_field}  and  Supplementary Material) and \(g = 27\), consistent with prior reports for flux-grown ZrTe\(_5\) samples exhibiting similar resistivity trends~\cite{sun2020large}. We can fit the oscillation extrema using only two free parameters, $v$ and $E_{\text{F}}$.  Fitting the oscillation extrema yields   \(v \approx 1.01(6) \times 10^5\) m/s and \(E_{\text{F}} \approx 6 (1)\) meV, in line with earlier studies~\cite{wang2018vanishing, tang2019three, liu2016zeeman, mutch2019evidence,wang2018discovery}.

Using these parameters, we compute the LL spectrum (Figure ~\ref{Landau_levels}b and \ref{Landau_levels}c). While we cannot definitively determine whether \(E_{\text{F}}\) lies in the conduction or valence band, the behavior of the LLs with field remains qualitatively the same in both cases. For illustrative purposes, we assume that \(E_{\text{F}}\) intersects the valence band to be consistent with experimental work on samples with similar electronic properties \cite{liang2018anomalous}. Despite the Fermi energy intersecting one of the electronic bands, the sample exhibits semiconducting behavior at higher temperature due to the low carrier concentration. This results from \(E_{\text{F}}\) residing close to the band edge, where thermal excitation across the gap remains the dominant transport mechanism. A supporting calculation included in the Supplementary Materials confirms that this scenario is consistent with the observed thermally activated resistance.

The evolution of the LLs with field provides further insight into the origin of the MR oscillations. At low magnetic fields (\(B < 3\)~T), the oscillations arise from spin-up and spin-down LLs within the valence band. At higher fields, however, the dominant contributions originate from spin-up LLs of the conduction band and spin-down LLs from the valence band. Consequently, in both low- and high-field regimes, spin-resolved LLs—offset in phase—govern the oscillatory behavior. Notably, the phase difference between these levels increases with field, consistent with the emergence of field-tunable interference discussed next. 

This behavior supports a scenario in which LL back-bending and spin-selective crossings drive the anomalous features observed in the quantum oscillation frequency.

\subsection{Landau level interference effect  in the temperature dependence of the oscillations}

\begin{figure}[H]
	\includegraphics[width=1\textwidth]{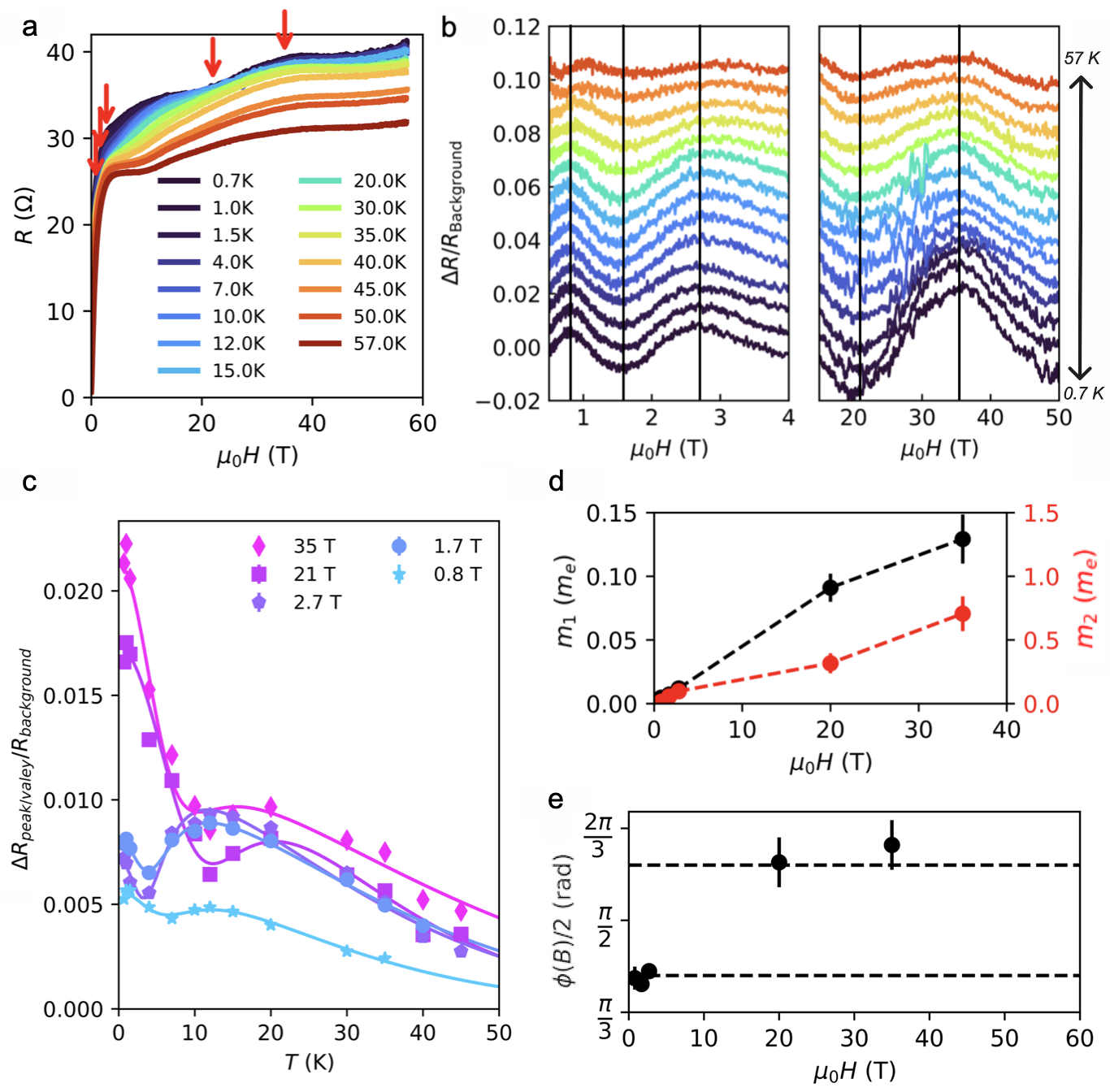}
	\centering
	\caption{ (a) Magnetoresistance of  ZrTe$_5$  for  various temperatures between 0.7 K-57 K. Measurements were done with same configuration illustrated in the inset of Figure  \ref{Pulsed_MR}a. The red arrows indicate the magnetic field where the temperature dependence of the $\Delta R_{\mathrm{peak/valley}} / R_{\mathrm{background}}$ was extracted. (b)  Oscillations as a function of magnetic field and temperature, the vertical black lines indicate the magnetic fields where the temperature dependence of the  $\Delta R_{\mathrm{peak/valley}} / R_{\mathrm{background}}$ were analyzed.  (c) $\Delta R_{\mathrm{peak/valley}} / R_{\mathrm{background}}$ at the magnetic field regimes indicated in (a) and (b). The dots represent experimentally obtained  $\Delta R_{\mathrm{peak/valley}} / R_{\mathrm{background}}$, while the continuous line depicts the extracted fitting using the electronic interference model described by Eq. \ref{Interference_model}.  (d)  Extracted electronic masses $m_1$ and $m_2$ related to spin-up and spin-down electrons as a function of magnetic field.  (e): The phase ($\phi$) between spin-up and spin-down crossings. The dashed lines indicates the average value of low ($\mu_0 H<$ 3T ) and high ($\mu_0 H>$20T ) field.}
    \label{Temperature_dependence_oscilations}
\end{figure}

Figure~\ref{Temperature_dependence_oscilations}a shows the temperature dependence of the MR of our ZrTe\(_5\) sample, measured from 700~mK to 60~K under magnetic fields up to 60~T. To isolate the MR oscillations, we subtracted a smooth background by fitting polynomial functions to the low- and high-field regions, specifically in the ranges [0.3~T, 5~T] and [15~T, 60~T], respectively. Due to the difficulty in fitting such a rapidly changing MR background, we restrict our analysis to the oscillation extrema marked by red arrows in Figure~\ref{Temperature_dependence_oscilations}a. In these locations the background remains smooth and the extracted oscillatory temperature dependence is robust. Within these intervals, we analyzed the temperature dependence of the MR oscillation for individual peaks and valleys at various magnetic fields. The peak and valley temperature dependence  was extracted by fitting a local sinusoidal function near each extreme; further details on the fitting and background subtraction are provided in the Supplementary Materials.

The temperature dependence of the oscillation amplitude, $\Delta R_{\mathrm{peak/valley}}/R_{background}$, directly reflects the temperature dependence of $A(B,T)$ for fixed field temperature cuts. Here, \(\Delta R_{\mathrm{peak/valley}}\) denotes the deviation of the resistance from the smooth background at each oscillation extremum (peak or valley)~\cite{pippard1965dynamics,mizzi2024reverse}. According to the LK model, the oscillation amplitude decreases monotonically with increasing temperature. In contrast, our data deviate from this behavior: a local minimum  in oscillation amplitude as a function of temperature consistently appears below 15~K at all magnetic fields (see Figure ~\ref{Temperature_dependence_oscilations}c).

To explain this anomaly, we consider an interference model in which two sets of LLs—originating from Zeeman-split spin-up and spin-down states—cross the Fermi energy with the same frequency \(F\) but with a phase offset~\cite{shoenberg2009magnetic, mccollam2005anomalous}. The total signal is modeled as \(\Delta R / R_\text{background} = a_{\uparrow}(B,T) \cos\left[2\pi\left(\frac{F(B)}{B} + \theta + \phi(B)\right)\right] + a_{\downarrow}(B,T) \cos\left[2\pi\left(\frac{F(B)}{B} + \theta\right)\right]\), where \(a_{\uparrow}\) and \(a_{\downarrow}\) are the amplitudes of the spin-up and spin-down contributions, respectively. The variable $\theta$ represents the intrinsic phase offset of the MR oscillations common to both spin components and \(\phi(B)\) is their relative phase.

This leads to an effective total amplitude \cite{mccollam2005anomalous,shoenberg2009magnetic}:
\begin{equation}
\begin{split}
A(T,B) = \Big[ &(a_{\uparrow}(T,B) + a_{\downarrow}(T,B))^2 
\cos^2\!\left(\frac{\phi(B)}{2}\right) \\
&+ (a_{\uparrow}(T,B) - a_{\downarrow}(T,B))^2 
\sin^2\!\left(\frac{\phi(B)}{2}\right) \Big]^{1/2}
\end{split}
\label{Interference_model}
\end{equation}

Each amplitude component follows an independent LK behavior, i.e., \(a_{\uparrow}(B,T) \propto \chi_{\uparrow} / \sinh(\chi_{\uparrow})\) and \(a_{\downarrow}(T,B) \propto \chi_{\downarrow} / \sinh(\chi_{\downarrow})\), where \(\chi_{\uparrow/\downarrow} = 2\pi^2 k_B T B / (\hbar e m_{\uparrow/\downarrow})\) and \(m_{\uparrow}\), \(m_{\downarrow}\) are the effective masses for the spin-up and spin-down branches, respectively. Eq. \ref{Interference_model} well fits the observed temperature dependence of $\Delta R_{\mathrm{peak/valley}}/R_{background}$ (solid lines, Figure \ref{Temperature_dependence_oscilations}c) for all isofield temperature cuts.

This explains not only the non-monotonic evolution of the oscillation amplitude versus temperature, but enables the extraction of the parameters of the two spin components contribution the interference model. While we denote them as spin-up and spin-down for convenience, the fitting alone does not allow us to assign the heavier or lighter mass to a specific spin orientation. For this reason, we label the extracted effective masses as \(m_1\) and \(m_2\), corresponding to the lighter and heavier branches, respectively. The extracted masses \(m_{1}\) and \(m_{2}\) are distinct and both increase with magnetic field (Figure ~\ref{Temperature_dependence_oscilations}d), indicating field-induced mass enhancement. The phase offset \(\phi\) also increases with field (Figure ~\ref{Temperature_dependence_oscilations}e), consistent with the LL calculation (Figure \ref{Landau_levels}b and c) where the phase difference between successive LL crossing increases with field.

This behavior may reflect spin-split bands with different curvatures, where Zeeman splitting yields one flatter (heavier-mass) and one steeper (lighter-mass) band. Alternatively, observed mass enhancement may arise from nonlinear LL evolution with magnetic field in the presence of Zeeman splitting, as seen in materials such as EuO and KTaO\(_3\)~\cite{rubi2023unconventional}. We discuss this in more detail in the Supplementary Material.

\subsection{3D-Fermi Surface}

AMR measurements  provide insight into anisotropic electronic transport by tracking resistance as a function of the angle between the applied magnetic field and crystallographic axes. In ZrTe\(_5\), we performed AMR by rotating the magnetic field within the \(bc\) and  \(ab\) planes (see  Supplementary Material). The angle \(\alpha\) is defined in the $bc$ plane (\(\alpha = 0^\circ\) along \(b\), \(90^\circ\) along \(c\)), and \(\beta\)  in the $ab$  (\(\beta = 0^\circ\) along \(b\), \(90^\circ\) along \(c\)).

\begin{figure}[H]
	\includegraphics[width=1\textwidth]{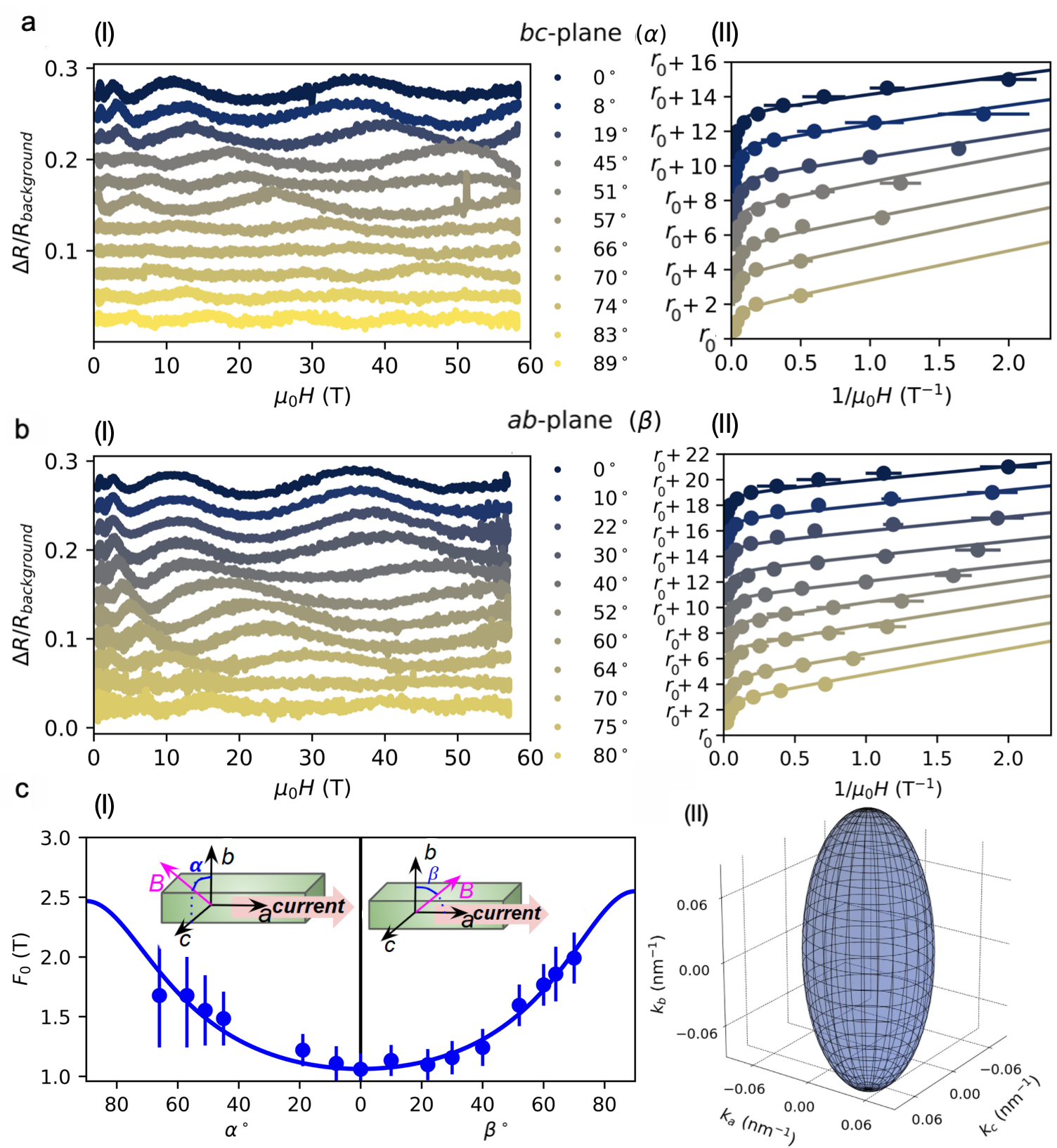}
	\centering
	\caption{    (a)  MR oscillations  measured at 4 K as a function of magnetic field up to 60 T for various angles \(\alpha\), defined in the \(bc\)-plane. (I)  MR after background subtraction. Curves are vertically offset for clarity. (II)  Magnetic field positions of MR peaks  and valleys, fit using Eq.~\ref{r-fitting}. (b) Same as (a), but with magnetic field rotated in the \(ab\)-plane by angle \(\beta\).  (c) (I) Extracted low field oscillation frequency \(F_0\) as a function of rotation angle across both planes. The error bars represent the fit uncertainties corresponding to a confidence interval of  $1\sigma$. The data follow a characteristic angular dependence of an ellipsoidal Fermi surface (solid line). The inset shows the schematic of the field and current directions relative to the crystallographic axes. (II) Schematic representation of an anisotropic Fermi surface with principal axes \(k_a\), \(k_b\), and \(k_c\) used to model the observed oscillation behavior.}
	\label{Angle_dependence}
\end{figure}

After proper background subtraction of the AMR, it is possible to observe that the MR oscillation  changes as a function of angle along the \(ab\) and \(bc\) planes, this is shown in panel I of Figure \ref{Angle_dependence} a and b.  We applied the same fitting procedure to the positions of the peaks and valleys of the oscillations as we did for the magnetic field along the \(b\), but now for all other angles. Fits to Eq. \ref{r-fitting}, are shown in panel II of Figure \ref{Angle_dependence} a and b. From this fit, it is possible to extract two relevant quantities: \(\frac{E_{\text{F}}^2 - M^2}{2e\hbar v^2}\) and \(\frac{\mu_B^2 g^2}{8e\hbar v^2}\), as a function of angle. The first of these has a particularly meaningful physical interpretation, as from Eq. \ref{Frequentcy}, in the zero field limit $F_0=\lim_{B \to 0}F(B) = \frac{E_{\text{F}}^2 - M^2}{2e\hbar v^2}$. This is the conventional $1/B$-periodic oscillatory contribution in the LK formula. By analyzing the angular dependence of \(F_0\), we aim to establish a connection between the $1/B$-oscillatory behavior observed in other samples \cite{kamm1985fermi,wang2018vanishing,tang2019three}  and the characteristics of our sample.

For ZrTe$_5$ samples that exhibit  a $1/B$ periodic MR, the extracted field independent QO frequency has angular dependence consistent with an ellipsoidal Fermi surface \cite{kamm1985fermi,tang2019three}.  The evolution of the QO frequency as a function of magnetic field angle for an ellipse can be described by the equation: 

\begin{equation}
F_{0}(\Theta)= \frac{F_0^bF_0^i}{\sqrt{(F_0^b \text{sin} (\Theta))^2+(F_0^i \text{cos} (\Theta))^2}}  \label{Frequency_angle_dependece}
\end{equation}

It is worth noting that for field-independent quantum oscillations,  Eq.~\ref{Interference_model} remains valid for frequencies extracted over any field interval.  In contrast, for our sample we define \(F_{0}^{\mathrm{b}}\) and \(F_{0}^{\mathrm{i}}\) as the zero-field limits  of the field-dependent oscillation frequencies, where \( F_0^b \) and \( F_0^i \) are the low field oscillation frequencies when the magnetic field is aligned along the \( b \) axis  and the in-plane axis \( i = a \) or \( c \), respectively. The angle \( \Theta \) denotes the tilt of the magnetic field away from the \( b \) axis , with \( \Theta = \alpha \) and \( \Theta = \beta \) corresponding to rotations in the \( bc \)- and \( ab \)-planes, respectively.

The angular dependence of $F_0$ extracted from our sample can be well-fitted by Eq. \ref{Frequency_angle_dependece}, as shown in Figure \ref{Angle_dependence}c panel I. From this fitting we extracted $F_0^b \approx 1.08(4)\,\text{T}$, $F_0^a \approx 2.6(5)\,\text{T}$, and $F_0^c \approx 2.4(5)\,\text{T}$, confirming the 3D  nature of the Fermi surface in our sample at low magnetic fields. This result suggests that the Fermi surface can be characterized by an ellipsoidal shape as illustrated in Figure \ref{Angle_dependence} c panel II. In this description, the Fermi wave-vectors are related to the oscillation frequencies by  $k_{F,i}=\big(\tfrac{2e}{\hbar}\tfrac{F_jF_k}{F_i}\big)^{1/2}$, with $(i,j,k)$ cyclic permutations of $(a,b,c)$. Using the measured frequencies, we obtain $k_a \approx 5.5(8)\times10^{-3}\,\text{\AA}^{-1}$, $k_b \approx 1.3(2)\times10^{-2}\,\text{\AA}^{-1}$, and $k_c \approx 6.0(9)\times10^{-3}\,\text{\AA}^{-1}$. This correspond to a Fermi-surface volume of $V_k \approx 1.8(5)\times10^{-6}\,\text{\AA}^{-3}$. The carrier density then follows as $n=\tfrac{g_s}{(2\pi)^3}V_k \approx 1.5(2)\times10^{16}\,\text{cm}^{-3}$,  where $g_s=2$ is the spin degeneracy. This further reinforces the anisotropic 3D electronic structure of our sample.  The density of carriers coincides with previous reported values for flux grown ZrTe$_5$ samples \cite{mutch2019evidence,wang2018discovery}.

The same 3D ellipsoidal characteristics are observed in samples with $1/B$-periodic oscillations. However, in these samples, the \(F_0\) values along the different axes are significantly larger, typically ranging from 1–5 T along the \(b\)  and 10–30 T along the other axes \cite{wang2018vanishing, tang2019three, galeski2021origin}. This suggests a larger Fermi surface and higher anisotropy across the different crystallographic directions. This can be attributed to the higher carrier density in CVT-grown samples, leading to an enhanced \(E_{\text{F}}\). The larger Fermi surface could explain why deviations from $1/B$-periodicity are not observed in these samples. Due to the higher \(E_{\text{F}}\), the \(F_0\) term dominates over the magnetic field correction in Eq. \ref{Frequentcy}, preventing observable deviations from the expected oscillatory behavior.

\section*{Discussion}

In this work, we have shown, through a combination of experimental data and theoretical modeling, that the origin of the non-$1/B$-periodic oscillations in ZrTe$_5$ lies in the reentrant LLs crossing the Fermi energy as a consequence of Zeeman-induced back-bending. The field dependence of the oscillation frequency is well fit by Eq. \ref{r-fitting},  directly reflecting the nonlinear evolution of the LLs in magnetic field. Interestingly, when the same data are plotted on a logarithmic field scale, the oscillations appear nearly log($B$)-periodic. This apparent scaling behavior emerges naturally from the analytic form of Eq.~\ref{r-fitting}. Expanding this expression around a characteristic field $B_0 = 20~\mathrm{T}$ yields $r(B) \simeq \phi + A\log(B) + A'[\log(B)]^2 + \cdots$, where $A = -\frac{E_F^2 - M^2}{2e\hbar v^2 B_0} - \frac{\mu_B^2 g^2 B_0}{8e\hbar v^2}$ and $A'=A/2$. Using the experimental parameters $E_F = 6~\mathrm{meV}$, $M = 5~\mathrm{meV}$, $v = 1.06\times10^5~\mathrm{m/s}$, and $g = 27$, we find $A \simeq -0.86$, in close agreement with the logarithmic slope extracted from the experimental fitting ($A\simeq-0.82$) shown in the inset of Figure ~\ref{Landau_levels}. This demonstrates that the apparent log($B$)-like spacing of oscillation extrema does not require invoking discrete-scale invariance but instead reflects the natural near-linear relation of $r(B)$ to $\log B$ arising from the interplay between the cyclotron and Zeeman terms in the reentrant regime.

In modeling the oscillations, we have adopted a fixed Fermi energy, a choice that is well justified for ZrTe$_5$ given the extremely low carrier density of its Dirac cones. In this limit, even a small electronic reservoir---arising from nearby parabolic bands or impurity states---can efficiently pin the Fermi level, preventing large field-induced shifts. Moreover, even in the clean limit where the chemical potential could, in principle, evolve with magnetic field, the back-bending of the LLs ensures that states continuously cross near the Fermi energy. This intrinsic reentrant behavior, will stabilize the Fermi energy around a nearly constant value, making the fixed-Fermi-energy approximation used in Eq.~\ref{r-fitting} physically consistent with the experimental behavior. Consistent with this picture, the oscillation extrema occur at nearly identical magnetic fields across all measured temperatures, while the background magnetoresistance shows only a weak temperature dependence. These observations confirm that variations in carrier mobility do not substantially influence the oscillation frequency or phase, and that temperature primarily affects the oscillation amplitude through the expected thermal damping resulting from Fermi-Dirac statistic and described by the LK formalism and interference effects.

\section{Conclusion} 

Our findings show the presence of non-$1/B$ MR oscillations up to high magnetic fields in ZrTe$_5$. These oscillations arise from the LL physics of a 3D Dirac electron system, where the competition between cyclotron energy and Zeeman splitting drives the LLs to backbend and re-cross the Fermi energy at higher magnetic fields. This same mechanism gives rise to interference effects between LLs, leading to clear deviations from the standard LK temperature dependence. Furthermore, the angle dependence of the zero-field oscillatory frequency follows an ellipsoidal model, consistent with a 3D Fermi surface and resembling the behavior of materials that exhibit conventional $1/B$-periodic oscillations. 

This framework reconciles the two regimes observed in ZrTe$_5$: the low-carrier-density regime, where non-$1/B$ oscillations arise and the field-dependent frequency induced by strong Zeeman splitting is apparent, and the high-carrier-density regime, where the larger Fermi energy masks this correction and restores the conventional $1/B$ behavior.  Our results identify reentrant LL recrossing the Fermi energy as the natural description of QO in ZrTe$_5$.

This work establishes ZrTe$_5$ as a realization of a 3D Dirac topological insulator, where all the unconventional features of the QOs are consistently captured by a minimal Dirac Hamiltonian. Remarkably, in this system the QL is reached in moderate magnetic fields ($<$3 T), yet the same Hamiltonian continues to describe the physics up to 60 T, including the field-driven enhancement of the effective mass that signals band flattening. Showing that such a simple description remains valid across this vast field range provides a robust framework for probing Dirac electrons under extreme conditions, leading to new paths to uncover novel instabilities and exotic quantum phases of matter.

\section{Methods}
\subsection{Sample Preparation} 

ZrTe$_5$ samples were grown using a self-flux method as described in reference \cite{mutch2019evidence}.   To improve the contacts on the sample,  gold contact pads were deposited by sputtering. Electrical contacts were made using 25 \(\mu\)m gold wire, which was fixed to the sample using highly conductive silver epoxy (EPO-TEC H20E).

\subsection{PPMS} 

The zero-field temperature dependence of the electrical resistance  (Figure \ref{ZrTe5_zero_field}) was measured using a four-point probe configuration using a 372- Resistance bridge from Lakeshore with an excitation current of of 31.6 $\mu$A. The sample were measured   down to 400 mK in a PPMS. 

\subsection{Pulsed Field Experiments}

MR measurements of our sample were conducted at the National High Magnetic Field Laboratory Pulsed Field Facility at Los Alamos National Laboratory. The sample was immersed in liquid helium (\( ^4\text{He} \)) for experiments conducted between 1.5 and 4 K, and in liquid (\( ^3\text{He} \)) for experiments below 1.5 K. For temperatures above 4 K, \( ^4\text{He} \) gas was used. The electrical resistance was measured using a four-point probe configuration. 

All measurements were performed using a lock-in technique with a frequency of 220kHz and current of 100 $\mu$A. A self-heating test was performed in zero field at 4 K and ohmic behavior was observed for an excitation current range of 10uA – 400uA. For the pulsed magnetic field measurements, the field profile in time is asymmetric with a rapid initial magnetic field rise to peak field and a comparatively slow magnetic field decay from peak field to zero. Eddy current heating was observed during the initial magnetic field rise, but the sample thermalizes with negligible eddy current heating during the magnetic field decay. Additional details are available in the Supplemental Information. All pulsed field data presented in this manuscript were taken from the falling magnetic field region of the magnetic field pulse.

 The sample was mounted on a 3D-printed cryogenic goniometer suitable for pulsed-field environments. In-situ rotation was performed with a precision of \( 0.2^\circ \), and the angular position was monitored using a mechanical counter and verified via pick-up coils mounted on the rotating stage \cite{Fedor_probe}.

\bmhead{Acknowledgments}

This work was performed at the National High Magnetic Field Laboratory, which is supported by the National Science Foundation through NSF/DMR-2128556, the State of Florida and the U.S. Department of Energy. The research presented in this article was supported by the Laboratory Directed Research and Development program of Los Alamos National Laboratory under project number 20230486ECR.  Material synthesis and characterization at the University of Washington was  supported by the Center on Programmable Quantum Materials, an Energy Frontier Research Center funded by the U.S. Department of Energy (DOE), Office of Science, Basic Energy Sciences (BES), under award DE-SC0019443. The views expressed in this article are those of the authors and do not reflect the official policy or position of the U.S. Naval Academy, Department of the Navy, the Department of Defense, or the U.S. Government. S. M. T. was supported by the Laboratory Directed Research and Development program of Los Alamos National Laboratory under project number 20230014DR. J. L. J acknowledges grants 2018/08845-3, 2019/24522-2 and 2022/15955-5, São Paulo Research Foundation (FAPESP) and CNPq 310065/2021-6. C. K. R acknowledges grant 2018/08845-3, 2019/24522-2, and 2022/15955-5 São Paulo Research Foundation (FAPESP) for support during the measurements. C.K.R., K.R., J.S., and N.H. and a portion of the data analysis and interpretation were supported by the U.S. Department of Energy “Science of 100 Tesla” BES program.


\bibliography{sn-bibliography}

\end{document}